\documentstyle[preprint,aps,epsf,floats,rotating]{revtex}

\newcommand{\beq}{\begin{equation}}
\newcommand{\eeq}{\end{equation}}
\newcommand{\bea}{\begin{eqnarray}}
\newcommand{\eea}{\end{eqnarray}}
\newcommand{\ben}{\begin{eqnarray*}}
\newcommand{\een}{\end{eqnarray*}}

\def\D0{D\O}

\newcommand{\epsl}{\not\!\epsilon}
\newcommand{\psl}{\not\!p}

\tightenlines

\begin{document}

\title{ Charm-Anticharm Asymmetries in Photoproduction \\
from Heavy-Quark Recombination}

\author{Eric Braaten, Yu Jia and Thomas Mehen\footnote{Address after Jan. 1, 2002:
Department of Physics, Duke University, Durham, NC 27708. }}

\address{Physics Department, Ohio State University, Columbus,
OH 43210, USA}

\date{\today}

\maketitle

\begin{abstract}

The asymmetries between charm and anticharm mesons
observed in fixed-target photoproduction experiments are
an order of magnitude larger than the asymmetries
predicted by conventional perturbative QCD.
We show that these charm meson asymmetries can be explained
by a heavy-quark recombination mechanism for heavy meson
production.  In this process, a charm quark combines with
a light antiquark from the hard-scattering process
and they subsequently hadronize into a state including
the charm meson.  This recombination mechanism can be
calculated within perturbative QCD up to some
nonperturbative constants. After using symmetries of QCD
to reduce the number of free parameters to two, we obtain a
good fit to all the data on the asymmetries for charmed
mesons from the E687 and E691 experiments.

\end{abstract}

\bigskip
\pacs{PACS number(s):  }

\thispagestyle{empty}

\newpage

\section{Charm-Anticharm asymmetries}

High energy charm photoproduction is an excellent testing ground for our
understanding of strong interactions\cite{Frixione:1998ma}.  The charm quark mass
provides a large scale justifying the use of perturbative QCD. Theoretical
uncertainties in photoproduction are smaller than in hadroproduction  because only
one hadron is present in the initial state. Charm events also constitute a higher
fraction of the total hadronic cross section than in  hadroproduction, which
facilitates experimental studies.

High energy charm photoproduction
experiments~\cite{Anjos:1989bz,Alvarez:1993yb,Frabetti:1996vi} have observed an
interesting production asymmetry between charm and anticharm mesons. The
E691~\cite{Anjos:1989bz} and E687~\cite{Frabetti:1996vi} experiments at Fermilab
measured the cross sections for various $D$ and $\overline{D}$ mesons from a photon
beam on a beryllium target. The experiments imposed a cut on the Feynman $x_F$
variable, $x_F>0$, which corresponds to the forward direction of the photon beam in
the $\gamma\,N$ center-of-mass frame. E691 and E687 found statistically significant
asymmetries in $D^+ - D^-$, $D^0 - \overline{D}^0$ and  $D^{*+} -
D^{*-}$. In each case  the cross section for the anticharm
meson is greater, though the size of the asymmetry is different for each species of
meson. Production asymmetries for $D^+_s$ and $\Lambda_c^+$ are also measured, but
with much larger errors so  they are consistent with zero. For $D^+$ and $D^{-}$
mesons, the E687 experiment has also  measured the dependence  of the charm
asymmetry on the photon energy, $E_\gamma$, the  meson transverse momentum,
$p_\perp$, and Feynman $x_F$.

The observed charm asymmetries in photoproduction are an order of magnitude larger
than the asymmetries between charm and anticharm quarks predicted by perturbative
QCD. Thus they cannot be understood within the fragmentation mechanism for heavy
meson production which is  based on the factorization theorem for inclusive single
particle production\cite{Curci:1980uw}. In this mechanism, the $c$ and
$\overline{c}$ are produced in a hard-scattering process and then  fragment
independently into charmed hadrons. At leading order in $\alpha_s$, charm is
produced by  photon-gluon fusion~\cite{Jones:1978wx} which produces $c$ and
$\overline c$ symmetrically.  The next-to-leading order (NLO)
correction~\cite{Ellis:1989sb,Frixione:1994dg,Smith:1992pw} increases the
normalization of the total cross section by about $50\%$ at the energies relevant
to the E687 and E691 experiments. The shapes of the $x_F$ and $p_\perp$
distributions are similar to those at leading order, but there is a tiny charm
asymmetry in these kinematic distributions, arising from the processes $\gamma +
q(\overline{q}) \rightarrow c + \overline{c} + q(\overline{q})$.
For $E_\gamma = 200$ GeV, the prediction from a NLO calculation
for the ratio of the $\overline{c}$ and $c$ cross sections
in the forward region is about $R=1.006$\cite{Cuautle:2000jf}.
The asymmetry, $(1-R)/(1+R)= -0.003$, is about an order of magnitude smaller than
the  asymmetries for $D$ mesons measured by the E687 and E691 experiments.
Furthermore, the NLO calculations cannot explain the differences in the asymmetries
of $D$ mesons with different light quark flavors. It is commonly assumed that a
nonperturbative hadronization mechanism involving the remnant of the nucleon or
photon after the hard scattering is responsible for the charm asymmetries.

In this paper, we argue that the charm asymmetries can be explained naturally and
economically by the heavy-quark recombination mechanism for heavy meson production
introduced in \cite{Braaten:2001bf}. In the $\overline{c}q$ recombination
mechanism,  a light $q$ from the target nucleon participates in the hard-scattering
process that produces the $c$ and $\overline{c}$. The light $q$ emerges from the
hard-scattering with small momentum in the $\overline{c}$ rest frame, and the
$\overline{c}$ and $q$ subsequently hadronize into a final state including the
$\overline{D}$ meson. There is also a $c\overline{q}$ recombination process in
which a light $\overline{q}$ from the target nucleon recombines with a $c$ to
hadronize into a $D$ meson. Because the parton distribution functions of $u$ and
$d$ in the nucleon are greater than the parton distribution functions for
$\overline{u}$ and $\overline{d}$, the recombination cross sections for
$\overline{c}u$ and $\overline{c}d$ are greater than for $c\overline{u}$ and
$c\overline{d}$. This gives rise to the asymmetries between charm and anticharm
mesons. Since the photon participates in the hard scattering and can couple to the
light quark, the recombination cross sections for processes involving $u$ quarks
differ from those involving $d$ and $s$ quarks because of their different electric
charges. Thus, the heavy-quark recombination mechanism can account for the
differences in the asymmetries of $D$ mesons with different light quark
flavor quantum numbers. Below
we will see that it can also adequately describe the
kinematic dependence of the asymmetries on $E_\gamma$, $p_\perp$ and $x_F$.

Before discussing the heavy-quark recombination mechanism in detail, we briefly
review the models which have been proposed in the literature to explain  the $D$
meson asymmetries. Though these models differ in details, all generate the
asymmetries by a nonperturbative hadronization process that involves the remnant of
the nucleon or photon after the hard scattering. The predictions of these models
depend on unknown and ad-hoc functions that specify the distribution of partons in
the remnant.

The model for the asymmetry that has been most thoroughly developed involves the
fragmentation of strings connecting the $c$ and $\overline{c}$ to the nucleon
remnant~\cite{Norrbin:2000zc}. After the $c$ and $\overline{c}$ are produced via
photon-gluon fusion, the remnant of the  target nucleon is left in a color-octet
state. The remnant is modelled as a color-triplet ``bachelor quark'' and a
color-antitriplet diquark with momentum fractions $x$ and $1-x$ specified by a
distribution function, $F(x)$.  The subsequent hadronization is calculated using
the Lund string fragmentation model~\cite{Sjostrand:1986ys}. Color-field strings
are introduced between the diquark and the $c$ and between the bachelor quark and
the $\overline c$. The prediction for the asymmetry depends sensitively on  the
choice of $F(x)$ as well as the relative probability for each string to break into a
charmed meson or a charmed baryon. This model reproduces the qualitative shapes of
the kinematic distributions for the $D^+-D^-$ asymmetry measured by the  E687
collaboration~\cite{Frabetti:1996vi}.  However, the magnitude of the asymmetries is
sensitive to the unknown function $F(x)$.  For a ``soft'' distribution that
diverges as $1/x$ at  small $x$ (the default option in the PYTHIA 5.6 event
generator program),  the model tends to overestimate the asymmetry in all
kinematic regions.  However, the data can be fit with a ``hard'' momentum
distribution proportional to $1-x$~\cite{Frabetti:1996vi}.

Models for the asymmetry that involve the remnant of the hadronic component of the
photon have also been proposed.  In one such model\cite{Cuautle:2000jf}, a
$\overline{c}$ is produced by the hard scattering process $q\overline{q}\rightarrow
c\overline{c}$ involving a $\overline{q}$ from the photon. The $\overline{c}$ then
recombines with a $q$ from the photon remnant to form a $\overline{D}$ meson. The
predictions depend on the distribution function for the $q$ in the photon remnant
and on a recombination function that could be calculated in principle using the
Lund string fragmentation model. Another model is based on the possibility of
intrinsic charm~\cite{Brodsky:1980pb} in the photon. In this model
\cite{Herrera:2001bg}, a $\overline{q}$ in the resolved photon undergoes a hard
scattering and the $\overline{D}$ forms from a $\overline{c}$ and $q$ in the photon
remnant. The prediction depends on the distribution function for the momentum
fractions of the $\overline{c}$ and $q$ in the photon remnant and a recombination
function that depends on several variables. Both of the models involve arbitrary
functions that can be tuned to fit the data.

In contrast with these models, the heavy-quark recombination mechanism
\cite{Braaten:2001bf} generates the asymmetries within a hard-scattering process.
The asymmetries are calculable in perturbative QCD up to some nonperturbative
parameters related to the probability for the $c$ and $\overline{q}$ produced in
the  short-distance process to hadronize into a $D$ meson. Because the $c$ and
$\overline{q}$ do not necessarily have the same color, angular momentum or light
quark flavor quantum numbers as the final state $D$ meson, several  nonperturbative
parameters appear in the calculation. However, it is possible to use heavy quark
symmetry, $SU(3)$ flavor symmetry and large $N_c$ arguments to reduce the number of
parameters to two. The asymmetries can be adequately described by fitting these two
parameters. Because of this, the heavy-quark recombination mechanism
gives a very economical description of the observed asymmetries
and does not rely on model-dependent
assumptions about nonperturbative hadronization.

\section{Heavy-quark recombination mechanism}

The factorization theorems of QCD \cite{Curci:1980uw} show that the leading
contribution to the $\overline{D}$ meson photoproduction cross section
at large transverse momentum can be
written as:
\beq    \label{fusion}
d\sigma[\gamma + N\rightarrow \overline{D} + X ]=  \sum_i
f_{i/N} \otimes
d\hat{\sigma}[\gamma + i \rightarrow c + \overline{c} + X] \otimes
D_{ \overline{c} \rightarrow \overline{D} }  \,,
\eeq
where $f_{i/N}$ is the parton distribution function for the parton $i$ in a nucleon
and $D_{ \overline{c} \rightarrow \overline{D} }$ is the nonperturbative
fragmentation function for $\overline{c}$ to hadronize into $\overline{D}$. At
leading order in $\alpha_s$, the only parton that contributes is the gluon, so
photon-gluon fusion dominates the cross section. At large $p_\perp$,
Eq.~(\ref{fusion}) receives corrections suppressed by powers of
$\Lambda_{QCD}/p_\perp$ or $m_c/p_\perp$. The heavy-quark recombination mechanism
introduced in \cite{Braaten:2001bf} is an $O(\Lambda_{QCD} m_c/p_\perp^2)$ power
correction to Eq.~(\ref{fusion}). The $\overline{c}q$ recombination contribution to
$\overline{D}$ production can be expressed as
\beq \label{bjm}
d\sigma[\gamma + N \rightarrow \overline{D} + X ] =
f_{q/N} \otimes
\sum_n d\hat{\sigma}[\gamma + q \rightarrow (\overline{c}q)^n +c]\:
\rho[(\overline{c}q)^n \rightarrow \overline{D} ]\, .
\eeq
Making the replacement $c \leftrightarrow \overline{c}$, $q \rightarrow
\overline{q}$ and $\overline{D}\rightarrow D$ in Eq.~(\ref{bjm}) gives the
$c\overline{q}$ recombination contribution to $D$ production. The notation
$(\overline{c}q)^n$ in Eq.~(\ref{bjm}) indicates that the $q$ in the final state
has momentum  of $O(\Lambda_{QCD})$ in the $\overline{c}$ rest frame, and that the
$\overline{c}$ and the $q$ are in a state with definite color and angular momentum
quantum numbers specified by $n$.  The factor $\rho[(\overline{c}q)^n \rightarrow
\overline{D}]$ is a nonperturbative constant proportional to the probability for
$(\overline{c}q)^n$ to evolve into a final state that includes the $\overline{D}$.
The recombination process in Eq.~(\ref{bjm}) is an $O(\Lambda_{QCD} m_c/p_\perp^2)$
power correction to Eq.~(\ref{fusion}), because the parton cross section in
Eq.~(\ref{bjm}) is suppressed by $O(m_c^2/p_\perp^2)$ relative to the parton cross
section in Eq.~(\ref{fusion}) and $\rho[(\overline{c}q)^n \rightarrow
\overline{D}]$ is expected to be $O(\Lambda_{QCD}/m_c)$ from heavy quark effective
theory scaling arguments.  Because the momentum of the light quark is
$O(\Lambda_{QCD})$, the production of $(\overline{c}q)^n$ in states with $L \neq 0$
is suppressed by powers of $\Lambda_{QCD}/p_\perp$ or $\Lambda_{QCD}/m_c$ relative
to S-waves.  Therefore, it is only necessary to compute production of
$(\overline{c}q)^n$ with  $^1S_0$ or $^3S_1$ angular momentum quantum numbers. The
$(\overline{c}q)^n$  can be in either a color-singlet or a color-octet state.

\begin{figure}[t]
  \centerline{\epsfysize=7.0truecm \epsfbox{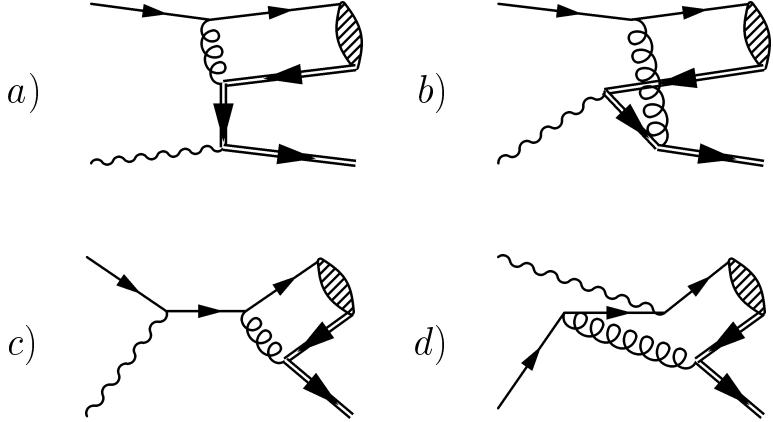}  }
 {\tighten
\caption{Feynman diagrams for the recombination process
$\gamma + q \rightarrow (\overline{c} q)^n + c$. Single lines represent a
light quark and double lines represents a charm quark. The shaded blob
represents $(\overline{c}q)^n$.}
\label{Feyn:Rec} }
\end{figure}

Berezhnoy, Kiselev, and Likhoded
\cite{Berezhnoi:1997st,Berezhnoy:1999rd,Berezhnoy:2000yj,Berezhnoy:2000ji}
have developed a model for heavy meson production
that incorporates some aspects of our heavy-quark recombination mechanism.
Their model is based upon an extrapolation of the NRQCD factorization
formalism for heavy quarkonium production \cite{Bodwin:1994jh}.
A $D$ or $D^*$ meson is modelled as a nonrelativistic quarkonium
composed of a $c$ and $\overline{q}$ with masses $m_c$ and $m_q$.
They take $m_q=300$ MeV,
consistent with an interpretation as a constituent quark mass.
Their expression for the cross section is obtained by keeping only specific
terms in the NRQCD factorization formula.
These terms involve S-wave matrix elements, which are distinguished by the
fact that their contributions at large $p_\perp$ are suppressed only by
$m_q m_c/p_\perp^2$.
Of the four independent S-wave matrix elements, they keep only two:
the Color-singlet Model matrix element
$\langle {\cal O}_1(^3S_1) \rangle^{D^*}
	= 3 \langle {\cal O}_1(^1S_0) \rangle^D$
and the color-octet matrix element
$\langle {\cal O}_8(^3S_1) \rangle^{D^*}
	= 3 \langle {\cal O}_8(^1S_0) \rangle^D$.
Thus their model involves only 3 parameters: $m_q$,
$\langle {\cal O}_1(^1S_0) \rangle^D$ and
$\langle {\cal O}_8(^1S_0) \rangle^D$.
In their model, the terms in the cross sections that diverge like
$1/m_q^2$ as $m_q \to 0$ have the same form as our heavy-quark
recombination cross sections if we make the following identification
between the nonperturbative factors:
\bea
\rho[c\bar q_1(^1S_0) \to D] &=&
{1 \over 4 m_q^2 m_c}\langle {\cal O}_1(^3S_1) \rangle^D \,,
\nonumber  \\
\rho[c\bar q_8(^1S_0) \to D] &=&
{1 \over 4 m_q^2 m_c}\langle {\cal O}_8(^3S_1) \rangle^D \,.
\eea
Our approach provides a model-independent framework
for calculating those contributions.

In the model of Ref.~\cite{Berezhnoy:2000ji},
contributions corresponding to the NRQCD matrix elements
$\langle {\cal O}_1(^3S_1) \rangle^D$ and
$\langle {\cal O}_8(^3S_1) \rangle^D$,
which require a spin-flip between the partons and the meson,
are omitted.  In the heavy quarkonium case, these matrix elements are
suppressed by $v^7$ and $v^3$ with respect to
$\langle {\cal O}_1(^1S_0) \rangle^D$, while
$\langle {\cal O}_8(^1S_0) \rangle^D$ is suppressed by $v^4$.
But when the antiquark $\bar q$ becomes light, $v$ approaches 1
and there is no parametric suppression
that can be used to justify omitting them.
Furthermore, the NRQCD operators ${\cal O}_1(^1S_0)$, ${\cal O}_1(^3S_1)$,
${\cal O}_8(^1S_0)$, and ${\cal O}_8(^3S_1)$ mix under renormalization.
This implies that at some order in perturbation theory,
inconsistencies will arise if any of these operators are omitted.
It is possible that some of them have matrix elements
that are numerically small, but they must all contribute at some level.

The most important difference between our approach
and the model of ref.~\cite{Berezhnoy:2000ji}
is that we use our nonperturbative
probability factors $\rho$ only for very specific terms
in the cross sections: those that diverge as $1/x_q$,
whre $x_q$ is the light-cone momentum fraction of the $\bar q$
relative to the heavy quark.
In contrast, the model of Ref.~\cite{Berezhnoy:2000ji}
is applied in all kinematical regions.  Their parameters $m_q$,
$\langle {\cal O}_1(^1S_0) \rangle^D$, and
$\langle {\cal O}_8(^1S_0) \rangle^D$
are used to calculate not only the heavy-quark recombination
contribution to the cross section, but all other contributions as well,
such as the fragmentation function $D_{c \to D}(z)$.
In contrast, we take the fragmentation function
to be an nonperturbative function that is independent of the
probability factors $\rho$ for heavy-quark recombination.
We use fragmentation functions which have been been
extracted from D meson production in $e^+e^-$ collisions.
With only 3 parameters,
the model of Ref.~\cite{Berezhnoy:2000ji} is very constrained,
but the predictive power comes at a sacrifice in rigor.

The four Feynman diagrams that contribute to the recombination process
$\gamma + q \rightarrow (\overline{c}q) + c$ at lowest order in $\alpha_s$
are shown in Fig.~\ref{Feyn:Rec}.
Since the momentum of the light quark in the final state,
$p_q^\prime$, is $O(\Lambda_{QCD})$ in the $\overline{c}$ rest frame, its
transverse momentum can be neglected and its light-cone momentum fraction,
$x_q$, is $O(\Lambda_{QCD}/m_c)$.  In the limit $x_q \rightarrow 0$,
the diagrams in Figs.~\ref{Feyn:Rec}a), b) and d) are $O(x_q^{-1})$
while the diagram in Fig.~\ref{Feyn:Rec}c) is $O(x_q^0)$.
Therefore the diagram in Fig.~\ref{Feyn:Rec}c)
gives an $O(\Lambda_{QCD}/m_c)$ suppressed contribution to the
heavy-quark recombination rate and can be neglected.
The calculation is very similar
to the calculation of the process $g + q \rightarrow (\overline{b} q) + b$
in Ref.~\cite{Braaten:2001bf}.
Interested readers can refer to \cite{Braaten:2001bf}
for a detailed description of the calculation.
The  color-singlet $^1S_0$ contribution to the cross section is obtained
by first making the following substitution in the parton amplitude:
\beq\label{sub}
v_i(p_{\overline{c}}) \overline{u}_j(p_q^\prime) \longrightarrow x_q
{\delta_{i j} \over N_c} m_c f_+ (\psl_{\overline{c}}-m_c)\gamma_5 \, .
\eeq
We then set $p_q^\prime = x_q p_{\overline{c}}$ in the rest of the amplitude and
take the limit $x_q \rightarrow 0$. The nonperturbative parameter $f_+$ is related
to the amplitude for the $(\overline{c}q)$ produced in the short-distance process
to bind into a $\overline{D}$ meson. $f_+$ is proportional to the $D$ meson  decay
constant, $f_D$, times a moment of the $\overline{D}$ meson light-cone wave
function, and therefore is expected to scale as $m_c^{-1/2}$ in the heavy quark
limit. However, $f_+$ does not account  for the possibility that the
$(\overline{c}q)$ can hadronize into states that include other light hadrons whose
momenta are soft in the $\overline{D}$ meson rest frame. In order to take these
final states into account, $f_+^2$ is replaced with the parameter
$\rho_1[\overline{c}q({^1S_0}) \rightarrow \overline{D}]$ after squaring the matrix
element. $\rho_1[\overline{c}q({^1S_0}) \rightarrow \overline{D}]$ is proportional
to the inclusive probability for the color-singlet, spin-singlet $(\overline{c}q)$
to evolve into any state containing the $\overline{D}$ hadron. The calculation of
the color-singlet $^3S_1$ cross section is identical except that $\gamma_5$ in
Eq.~(\ref{sub}) is replaced by $\epsl$, where $\epsilon^\mu$ is a polarization
vector, and $f_+^2$ is replaced by $\rho_1[\overline{c}q({^3S_1}) \rightarrow
\overline{D}]$.

The $\overline{D}$ production cross section also receives contributions from
processes in which the $(\overline{c}q)$ is produced in a color-octet configuration
in the short-distance process. The color-octet $^1S_0$ cross section is obtained by
making the substitution
\bea\label{sub8}
v_i(p_{\overline{c}}) \overline{u}_j(p_q^\prime) \longrightarrow x_q
\sqrt{2\over N_c} T^a_{i j}\,m_c f_+^8(\psl_{\overline{c}} - m_c)\gamma_5 \,
\eea
in the parton amplitude, replacing $p_q^\prime \rightarrow x_q p_{\overline{c}}$
and then taking the limit $x_q \rightarrow 0$. After squaring the matrix element
and summing over colors in the final state, $(f_+^8)^2$ is replaced with
$\rho_8[\overline{c}q ({^1S_0}) \rightarrow \overline{D}]$. The normalization in
Eq.~(\ref{sub8}) is chosen so that $\rho_8/\rho_1$ is the ratio of the
probabilities for a color-octet $(\overline{c}q)$ and a color-singlet
$(\overline{c}q)$ to hadronize into a $D$ hadron. Finally the color-octet  $^3S_1$
contribution  is obtained by replacing $\gamma_5$ in Eq.~(\ref{sub8})  by $\epsl$,
and replacing $(f_+^8)^2$ by  $\rho_8[\overline{c}q ({^3S_1}) \rightarrow
\overline{D}]$.

The final results for the parton cross sections for
color-singlet $\overline{c}q$ recombination in
$\gamma + q$ collisions are
\bea\label{ss}
  {d\hat{\sigma} \over d\hat{t}}[\overline{c}q(^1S_0^{(1)})] &=&
\,{256\,\pi^2\,e_c^2\,\alpha\,\alpha_s^2 \over 81}  {m_c^2 \over S^3}
\left[ - {S\over U} \left ( 1+ {\kappa\,T \over S} \right)^2   \right.
\nonumber  \\
& &   \left.   + \,{m_c^2\,S \over U^2}\left( -{S^3 \over T^3} +
{ 2\,(1+ \kappa) S \over T }  + 4\,\kappa +  {\kappa^2\,T \over S} \right)    +
\, {2\,m_c^4\,S^3 \over T^3\,U^2}\left(1+ {\kappa\,T\over S} \right) \right] \,,
 \\ \label{st}
{d \hat{\sigma} \over d\hat{t}}[\overline{c}q(^3S_1^{(1)})] &=&
\,{256\,\pi^2\,e_c^2\,\alpha\,\alpha_s^2 \over 81}  {m_c^2 \over S^3} \left[
- {S\over U} \left( 1 + {2\,U^2 \over T^2 } \right ) \left (1+
{\kappa\,T \over S} \right)^2  \right.  \nonumber \\
& &  +\, {m_c^2\,S \over
U^2}\left( {S^3 \over T^3} + {4\,(2+ \kappa )\,S^2 \over T^2} + {2\,(3+ 7\,\kappa
)\,S \over T} + 4\,\kappa \,( 3+ \kappa) +{ 3\,\kappa^2\,T \over S}  \right )
\nonumber \\
& & \left.  + \, {6\,m_c^4 \,S^3 \over T^3\,U^2} \left(1+ {\kappa
\,T\over S} \right)\right] \, ,
\eea
where $\kappa = e_q/e_c$ is the ratio of the electric charge of the light quark $q$
to that of the charm quark, $S=(p_{q}+p_\gamma)^2$, $T=(p_\gamma-p_c)^2-m_c^2$ and
$U = (p_\gamma-p_{\overline{c}})^2-m_c^2$. These variables are defined so that
$S+T+U=0$. If $\theta$ is defined as the angle between the incoming $q$ and the
outgoing $(\overline{c}q)$ in the parton center-of-mass frame, then $T
=-S/2[1-(1-4 m_c^2/S)^{1/2}\cos\theta]$. The contributions to the cross section for
$\overline{D}$ meson production are obtained by multiplying the parton cross
sections in Eqs.~(\ref{ss}) and (\ref{st}) by
$\rho_1[\overline{c}q(^1S_0)\rightarrow \overline{D}]$ and
$\rho_1[\overline{c}q(^3S_1)\rightarrow \overline{D}]$, respectively. Note that the
recombination process is calculated in the heavy quark limit, so we do not
distinguish between $m_c$ and the $\overline{D}$ meson mass.

For the photoproduction diagrams in Fig.~\ref{Feyn:Rec}, the color-octet amplitude
is identical to the color-singlet amplitude up to an overall color factor. The
color-octet cross sections can be obtained from the color-singlet
cross sections in Eqs.~(\ref{ss}) and (\ref{st})
by replacing $\rho_1$ with $\rho_8/8$.
Therefore, heavy-quark recombination contributions to
photoproduction  cross sections depend on the following linear combination of
color-singlet and  color-octet parameters:
\beq
\rho_{\rm eff} = \rho_1 + {1\over 8}\,\rho_8\,.
\eeq

It is interesting to compare the relative sizes of the heavy-quark recombination
and photon-gluon fusion cross sections in different regions of phase space.
Setting $\theta = \pi/2$ and expanding in powers of $m_c^2/S$, one finds that the
ratio of the parton cross sections is
\bea
\left.
{d \hat{\sigma}[\gamma + q \rightarrow (\overline{c}q)+c]  \over
d \hat{\sigma}[\gamma + g\rightarrow \overline{c}+c]}\right|_{\theta = \pi/2}
&\approx &\, {64\,\pi\, (2-\kappa)^2 \over 81} \alpha_s {m_c^2 \over S}
\qquad {\rm for } \;\,\, \overline{c}q(^1S_0) \, , \\
&\approx &\, {64\,\pi\,(2-\kappa)^2 \over 27} \alpha_s{m_c^2 \over S}
\qquad {\rm for} \;\,\, \overline{c}q(^3S_1) \, .
\eea
For $\theta = \pi/2$, $S =4(p_\perp^2+m_c^2)$, so the parton cross section for
heavy-quark recombination is suppressed by $m_c^2/p_\perp^2$ at large $p_\perp$. The
recombination cross sections are therefore suppressed by
$\Lambda_{QCD}m_c/p_\perp^2$ relative to the photon-gluon fusion cross section, in
accord with the factorization theorem. The differential cross section is finite  as
$\theta \rightarrow 0$ and $\theta \rightarrow \pi$ because the heavy quark mass
acts as an infrared cutoff. Setting  $\theta=0$ and expanding in $m_c^2/S$,
one finds that the ratio of the two parton cross sections is
\bea \label{ratio}
\left.
{d \hat{\sigma}[\gamma + q\rightarrow (\overline{c}q)+c]\over
d \hat{\sigma}[\gamma + g\rightarrow \overline{c}+c]}\right|_{\theta = 0}
\approx {256 \,\pi \over 81} \alpha_s
\qquad \qquad{\rm for}\;\,\, \overline{c}q(^1S_0) \, , \overline{c}q(^3S_1) \, .
\eea
Thus, the heavy-quark recombination cross sections
have no kinematic suppression factor when
the $(\overline{c}q)$ emerges from the parton collision in the same direction as
the incident $q$. On the other hand, the recombination cross sections are suppressed
by factors of $m_c^2/S$ when the $(\overline{c}q)$ emerges in the same direction as
the incident photon:
\bea \label{backsct}
\left. {d \hat{\sigma}[\gamma + q\rightarrow (\overline{c}q)+c] \over
d \hat{\sigma}[\gamma + g\rightarrow \overline{c}+c]}\right|_{\theta = \pi}
&\approx & {256 \,\pi \over 81} \alpha_s {m_c^6\over S^3}\,
\qquad \qquad \quad \,\,\,\,{\rm for}\,\,\, \overline{c}q(^1S_0) , \\
&\approx &{512 \,\pi \,(1+\kappa^2)\over 81}\alpha_s {m_c^2\over S} \,
\qquad {\rm for} \,\,\, \overline{c}q(^3S_1) \, .
\eea
The kinematic suppression is especially strong in the $^1S_0$ case.

To analyze the photoproduction data of E687 and E691, it is necessary to compute
the  cross section for various species of $D$ mesons. It is important to realize
that the color, angular momentum and light quark flavor quantum numbers of the $(c
\overline{q})^n$ are not necessarily the same as the $D$ meson in the final
state\cite{Braaten:2001bf}. The transition from the $(c\overline{q})^n$ to the
final state that includes the $D$ meson is governed by nonperturbative physics
that can change these quantum numbers. For instance it is possible for the
$(c\overline{q})$ to be produced in a color-octet state in the short-distance
process and emit soft gluons during the nonperturbative  transition to emerge as a
color-singlet $D$ meson.  It is also possible for the spin of the light quark to be
flipped during this transition, resulting in a $D$ meson  with different angular
momentum. Finally, a light $q^\prime \overline{q}^\prime$ pair can be created in
the transition, followed by the $c$ binding with the $\overline{q}^\prime$ to form
a $D$ meson with different flavor quantum numbers than the original
$(c\overline{q})^n$. Thus for a specific $D$ meson, several nonperturbative
parameters enter the calculation. For instance, for the $D^0$ meson, the parameters
are $\rho_{\rm eff}[c\overline{u}(^1S_0) \rightarrow D^0]$ and $\rho_{\rm
eff}[c\overline{u}(^3S_1) \rightarrow D^0]$, together with those obtained by
replacing $\overline{u}$ with $\overline{d}$ or $\overline{s}$.

Symmetries of the strong interaction allow us to greatly reduce the number of free
parameters. Heavy quark spin symmetry \cite{Isgur:1989vq} implies that the spin of
the $c$ quark should remain unchanged throughout the nonperturbative transtion.
This means that transitions which can be related by flipping the spin of the heavy
quark in both the initial and final state should have the same rate. For example,
\bea\label{hqs}
\rho_{\rm eff}[c\overline{d}(^1S_0) \rightarrow D^0] =\rho_{\rm eff}[c\overline{d}(^3S_1)
\rightarrow D^{*0}] \, .
\eea
One can also use $SU(3)$ light quark flavor symmetry to relate nonperturbative
transitions with different light quark flavors. An example of an $SU(3)$  relation
is
\bea\label{su3}
\rho_{\rm eff}[c\overline{u}(^1S_0) \rightarrow D^0] =
\rho_{\rm eff}[c\overline{d}(^1S_0) \rightarrow D^{+}] =
\rho_{\rm eff}[c\overline{s}(^1S_0)\rightarrow D_s^{+}] \, .
\eea
Finally, one can argue that in the large $N_c$ limit of QCD, light quark-antiquark
pair production is suppressed. This suppression in low energy QCD is the basis for
the phenomenologically successful Zweig's rule.  Suppression of light
quark-antiquark pair production in the  long-distance transition then forces the
$(c\overline{q})^n$ to have the same flavor quantum numbers as the final state $D$
meson.

Using the symmetries discussed above
together with the large $N_c$ argument,
one is left with two parameters. One is
\bea\label{sm}
 \rho_{\rm sm} &=& \rho_{\rm eff}[c\overline{u}(^1S_0) \rightarrow D^0]
 = \rho_{\rm eff}[c\overline{d}(^1S_0) \rightarrow D^+]
 = \rho_{\rm eff}[c\overline{s}(^1S_0) \rightarrow D_s^+] \, \nonumber \\
&=&  \rho_{\rm eff}[c\overline{u}(^3S_1) \rightarrow D^{*0}]
 = \rho_{\rm eff}[c\overline{d}(^3S_1) \rightarrow D^{*+}]
 = \rho_{\rm eff}[c\overline{s}(^3S_1) \rightarrow D_s^{*+}] \, .
\eea
The subscript sm stands for spin-matched, since in all these transitions, the
spin of the $c\overline{q}$ is the same as the spin of the $D$ meson.
The other parameter describes transitions in which the light quark has its spin
flipped:
\bea \label{sf}
 \rho_{\rm sf} &=& \rho_{\rm eff}[c\overline{u}(^3S_1) \rightarrow D^0]
 = \rho_{\rm eff}[c\overline{d}(^3S_1) \rightarrow D^+]
 = \rho_{\rm eff}[c\overline{s}(^3S_1) \rightarrow D_s^+] \, \nonumber \\
&=& \rho_{\rm eff}[c\overline{u}(^1S_0) \rightarrow D^{*0}]
 = \rho_{\rm eff}[c\overline{d}(^1S_0) \rightarrow D^{*+}]
 = \rho_{\rm eff}[c\overline{s}(^1S_0) \rightarrow D_s^{*+}]  \, .
\eea
While the use of heavy quark and $SU(3)$ symmetry greatly reduces the number of
parameters, it should be kept in mind that these relations are only expected to
hold  at the $30\%$ level. It is also important to note that the flavor-changing
transitions are not strongly suppressed because $N_c =3$.  The relevant
nonperturbative parameters for these transitions are expected  to be smaller than
$\rho_{\rm sm}$ and $\rho_{\rm sf}$ but nonvanishing. These are not included in the
calculations of this paper because an adequate fit to E687 and E691 data can be
obtained using only the two parameters $\rho_{\rm sm}$ and $\rho_{\rm sf}$.
However, it may be necessary to include flavor-changing transitions in other
applications, or if the accuracy of current data on charm asymmetries significantly
improves.

It is important to note that when a $\overline{D}$ is produced by a
$\overline{c}q$ recombination process, the associated $c$ quark can produce a $D$
meson via the usual fragmentation mechanism. Therefore, the recombination process
can contribute to the direct cross section for $D$ mesons in two ways:
\bea\label{direct}
&a)& \qquad \sum_n d\hat{\sigma}[\gamma + \overline{q} \rightarrow
(c\overline{q})^n + \bar{c}] \,
\rho_n [(c\overline{q})^n \rightarrow D ] \, ,
\\ \label{backside}
&b)& \qquad \sum_{q,\,n}d\hat{\sigma}[\gamma + q \rightarrow
(\overline{c} q)^n+ c]\, \rho_n \otimes D_{c\rightarrow D} \, .
\eea
In $a)$, the $(c\overline{q})^n$ recombines into the $D$ meson. In $b)$, the
$(\overline{c}q)^n$ recombines into a $\overline{D}$ meson, and the $D$ comes from
the fragmentation of the recoiling $c$.  For direct $\overline{D}$ production,
Eq.~(\ref{direct}) and Eq.~(\ref{backside}) are replaced with their charge
conjugates. Process $a)$ produces more $D^-$ and $\overline{D}^0$ mesons than $D^+$
and $D^0$, simply because the nucleon contains more $u$ and $d$ than $\overline{u}$
and $\overline{d}$.  In process $b)$, $D$'s are produced when there is a $q$ in the
initial state, while $\overline{D}$'s are produced when there is a $\overline{q}$
in the initial state. The excess of $u$ and $d$ over $\overline{u}$ and
$\overline{d}$ in the nucleon then leads to more $D$ than $\overline{D}$. It turns
out that in the case of $D^+-D^-$ and $D^0-\overline{D}^0$ mesons, process $b)$
dilutes the asymmetry produced in $a)$. For $D_s$ mesons, process $a)$ cannot
generate an asymmetry because the $s$ and the $\overline{s}$ content of the nucleon
are identical. Process $b)$ generates an asymmetry for $D_s^+-D_s^-$ which has the
opposite sign as the $D^+-D^-$ and $D^0-\overline{D}^0$ asymmetries.

Finally, when computing the inclusive cross section for $D$ and $D^*$ mesons, one
must consider the possibility of feeddown from other $D$ mesons that are produced
via photon-gluon fusion or heavy-quark recombination
and then decay to $D$ or $D^*$. The
feeddown from excited $D$ mesons, e.g. $D^*_2$, is not included but is expected to
be small.  Therefore, the inclusive $D^*$ cross section is the sum of the photon-gluon
fusion and heavy-quark recombination contributions, which will be referred to as the direct
cross section. The inclusive $D$ meson cross section also includes feeddown from
$D^*$ mesons. Denoting the direct cross sections by $\sigma_{\rm dir}$  and the
inclusive cross sections by $\sigma_{\rm inc}$, one finds
\bea \label{feeddown}
\sigma_{\rm inc} [D^+] & = & \sigma_{\rm dir}[D^+]+ 0.323\,
\sigma_{\rm dir} [D^{*+}]\,,
\label{D+} \\
\sigma_{\rm inc} [D^0] & = & \sigma_{\rm dir}[D^0]+ 0.677\,
\sigma_{\rm dir} [D^{*+}] +  \sigma_{\rm dir} [D^{*0}]\,,
\label{D0} \\
\sigma_{\rm inc} [D_s^+] & = & \sigma_{\rm dir} [D_s^+] +
\sigma_{\rm dir}[D_s^{*+}]\,.
\eea
The coefficients of $\sigma_{\rm dir}[D^{*+}]$ in Eq.~(\ref{D+}) and Eq.~(\ref{D0})
are the branching fractions of $D^{*+}$ into $D^+$ and $D^0$~\cite{Groom:2000in}.
The cross sections for the charge conjugate states have similar expressions.

\section{Comparison with photoproduction data}

In the calculations presented in this paper, the contribution to charm
photoproduction from the hadronic component of the photon is neglected because it
constitutes less than $5\%$ of the cross section for the fixed-target experiments
in the energy range $50 \;{\rm GeV}< E_\gamma < 400 \;{\rm
GeV}$\cite{Ellis:1989sb}. The charm quark mass is taken to be  $1.5\:{\rm GeV}$ and
both the factorization and renormalization scales are set equal to
$\sqrt{m_c^2+p_\perp^2}$. We checked that varying this scale between
$\sqrt{m_c^2+p_\perp^2}/2$ and $2\sqrt{m_c^2+p_\perp^2}$ has a small effect on the
predicted asymmetries. The CTEQ5L \cite{Lai:2000wy}  parton distribution functions
are used. For this parton set, it is  appropriate to use the one-loop running
strong coupling constant with four active flavors and $\Lambda_{QCD} = 0.192\:{\rm
GeV}$.  Since the E691 and E687 collaborations used a beryllium target, the
appropriate parton distribution is a linear combination of proton and neutron
parton distribution functions: $f_{i/Be}=(4/9)f_{i/p} + (5/9)f_{i/n}$.

We use the following fragmentation function for the $D$ meson:
\beq
  D_{c\rightarrow D} (z) = D(z;\epsilon) f_{c \rightarrow D}\,,
\eeq
where $f_{c\rightarrow D}$ is the probability for the direct fragmentation of $c$
into $D$ and $D(z;\epsilon)$ is the Peterson fragmentation
function\cite{Peterson:1983ak}, normalized so that $\int_0^1 \,D(z;\epsilon) dz
=1$.  Recent OPAL measurements find $f_{c\rightarrow D^{*+}} = 0.22 \pm
0.02$~\cite{Ackerstaff:1998ki} and $f_{c\rightarrow D^{*0}} = 0.22 \pm
0.07$~\cite{Ackerstaff:1998as}. Heavy-quark spin symmetry implies that the direct
fragmentation probabilities for $D^+$ and $D^0$ are smaller by a factor of 3.
The probability for fragmentation to $D_s^+$ is smaller
than the probability for $D^0$ and $D^+$ by
roughly a factor of two \cite{Frixione:1998ma}.
In this paper, the fragmentation
probabilities are chosen to be
\bea\label{frag}
f_{c\rightarrow D^{*+}} & = & 3\,f_{c\rightarrow D^+} = 0.22\,,
\\
f_{c\rightarrow D^{*0}} & = & 3\,f_{c\rightarrow D^0} = 0.22\,,
\\
f_{c\rightarrow D_s^{*+}} & = & 3\,f_{c\rightarrow D_s^+} = 0.11 \,.
\eea
These direct fragmentation probabilities add up to about $73\%$, and
the remaining probability comes from fragmentation to charmed baryons.

Fragmentation softens the $p_\perp$ and $x_F$ distributions of the $D$ mesons
relative to the $c$ quarks produced in the hard scattering process.
The measured value of the Peterson parameter $\epsilon$ depends
on the order to which the perturbative factors have been calculated.
The measured values of $\epsilon$ corresponding to leading order
perturbative calculations range from $0.04$ to $0.13$
depending on the $D$ hadron \cite{Groom:2000in}.  It is worth noting that the
measured values of $\epsilon$ for $D$ mesons tend to be larger than for $D^*$
mesons. Larger $\epsilon$ corresponds to a softer $D$ meson distribution. The
softening of the $D$ meson  fragmentation function is due primarily to the feeddown
contributions from $D^*$ decay. For simplicity, we neglect the softening of the $D$
meson fragmentation function and use a common value $\epsilon = 0.06$ for all $D$
and $D^*$ mesons. Varying $\epsilon$ between  0.04 and 0.13 has a minimal effect
on  the predictions for the asymmetries.

\begin{figure}[!t]
 \centerline{\epsfysize=12.0truecm \begin{turn}{270} \epsfbox{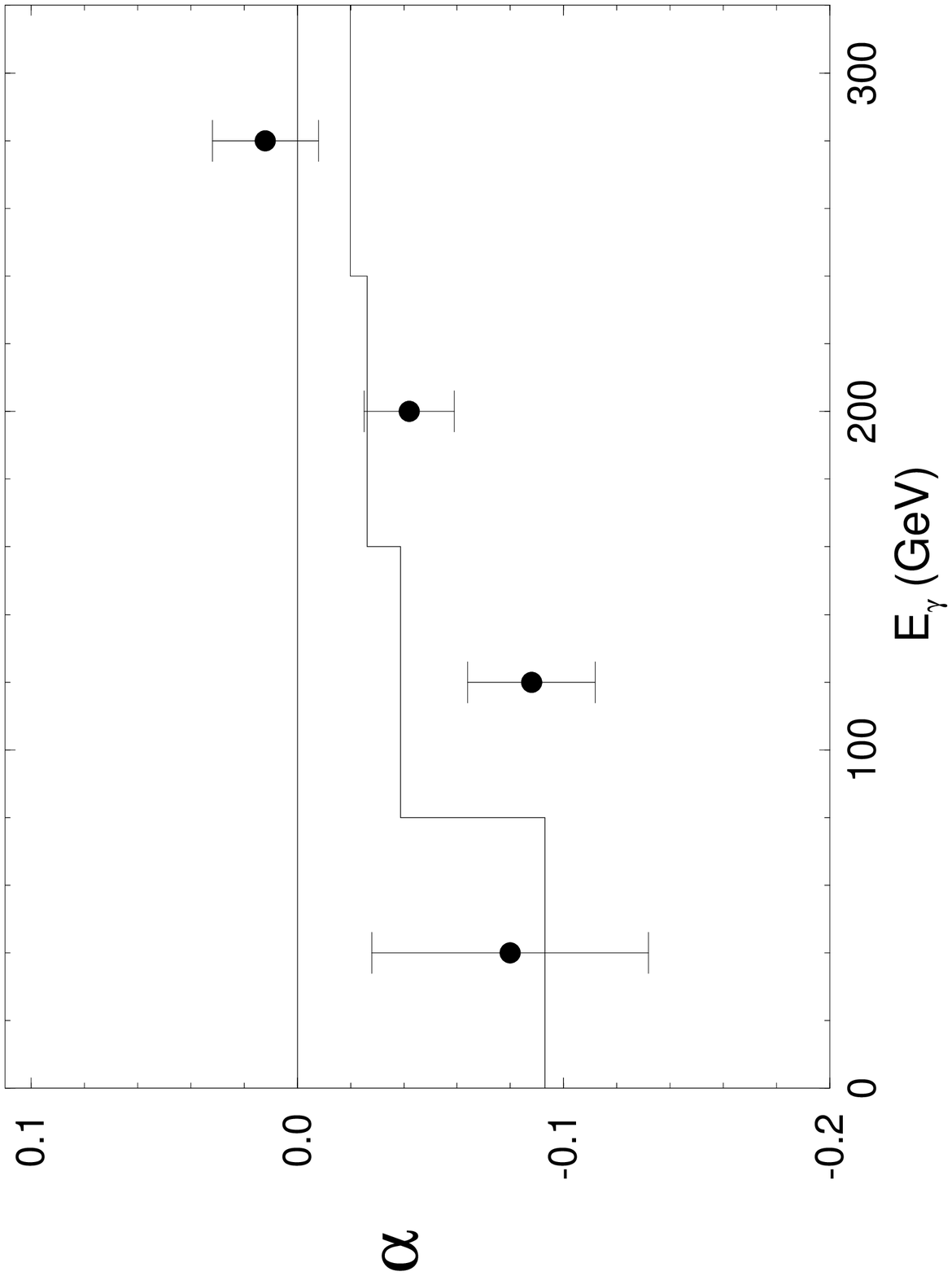}
  \end{turn} }
 \vspace{0.25 in}
 {\tighten
\caption[1]{
The $E_\gamma$ dependence of $\alpha[D^+]$. The data points
are the measurements of E687 and the histogram is our
prediction with $\rho_{\rm sm} = 0.15$, $\rho_{\rm sf} = 0$.}
\label{Eg}}
\end{figure}

\begin{figure}[!t]
  \centerline{\epsfysize=12.0truecm \begin{turn}{270} \epsfbox{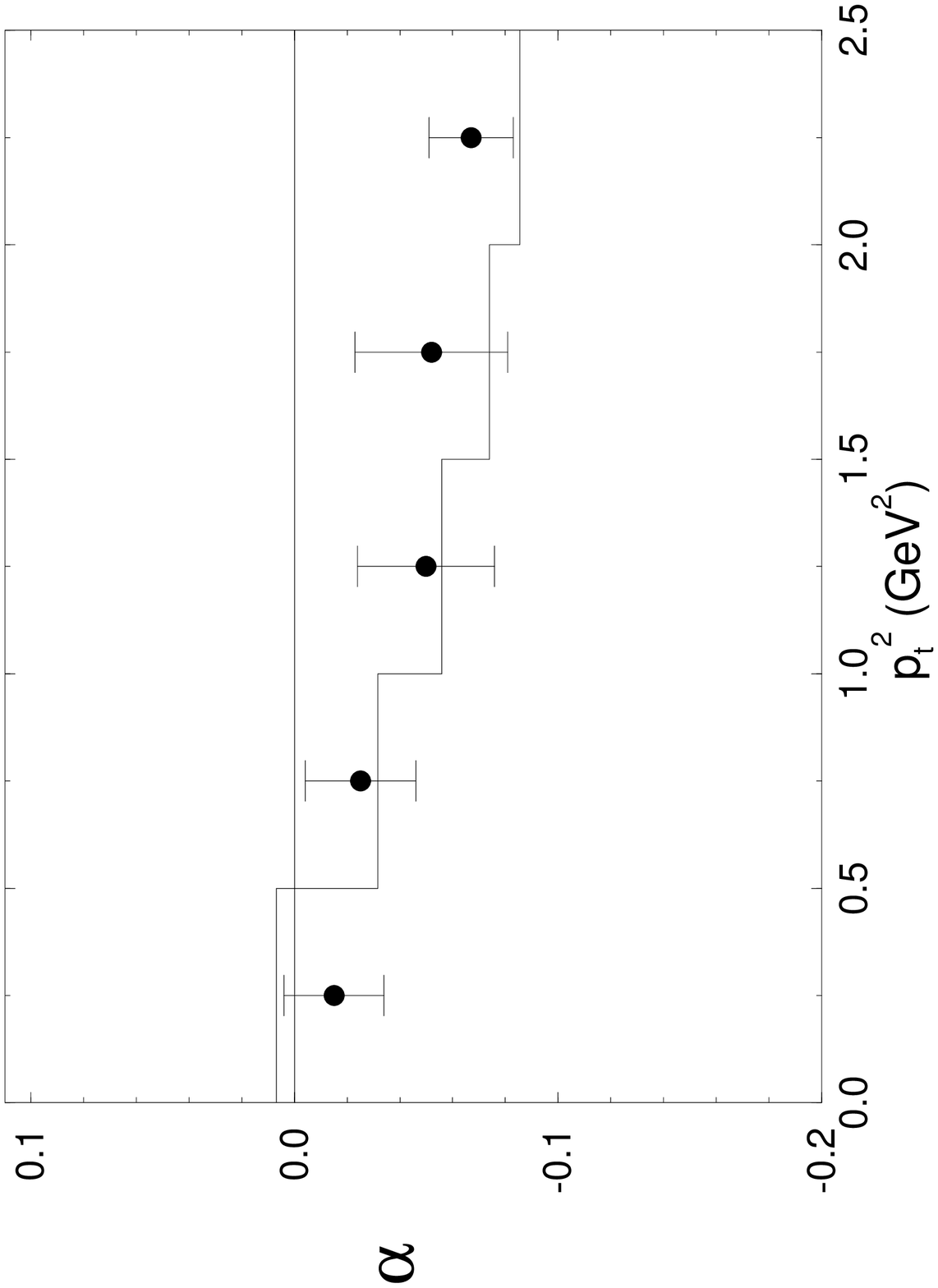}
 \end{turn} }
 \vspace{0.25 in}
 {\tighten
\caption[1]{
The $p_\perp^2$ distribution of $\alpha[D^+]$. The data points
are the measurements of E687 and the histogram is our
prediction with $\rho_{\rm sm} = 0.15$, $\rho_{\rm sf} = 0$. }
\label{ptsqr} }
\end{figure}

\begin{figure}[!t]
  \centerline{\epsfysize=12.0truecm
  \begin{turn}{270} \epsfbox{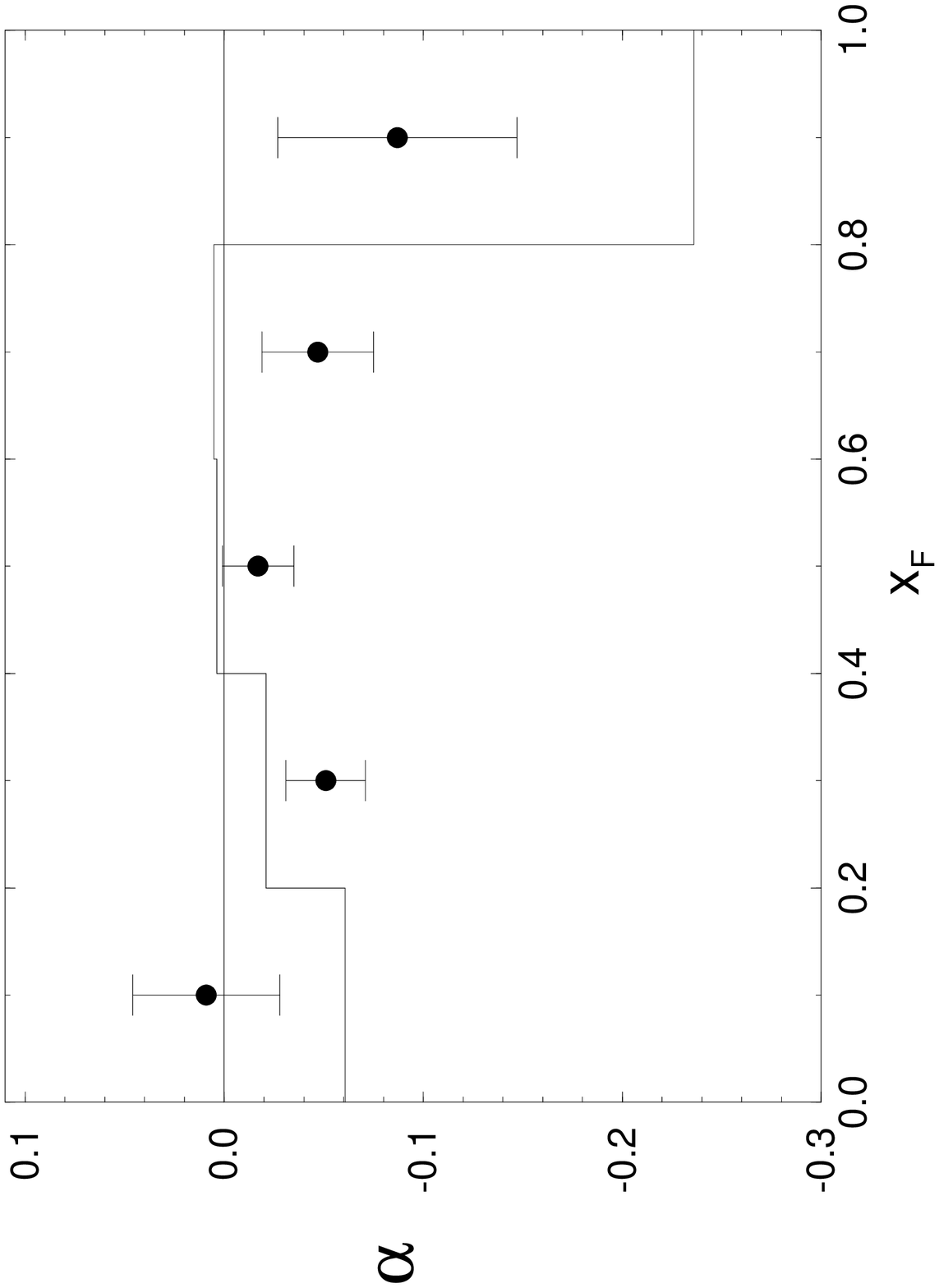}\end{turn}  }
   \vspace{0.25 in}
  {\tighten
\caption[1]{
The $x_F$ distribution of $\alpha[D^+]$. The data points
are the measurements of E687 and the histogram is our
prediction with $\rho_{\rm sm} = 0.15$, $\rho_{\rm sf} = 0$.}
\label{xf}}
\end{figure}

Measurements of the charge asymmetry for a specific $D$ meson
are usually expressed either in terms of the anticharm/charm production ratio
defined by
\bea
R[D]={\sigma_{\rm inc}[\overline D] \over \sigma_{\rm inc}[D]}\, ,
\eea
or in terms of the asymmetry variable
\beq
\alpha[D] = { \sigma_{\rm inc}[D] -\sigma_{\rm inc}[\overline{D}]  \over
           \sigma_{\rm inc}[D] + \sigma_{\rm inc}[\overline{D}]  } \, .
\eeq
The E687 and E691 measurements of $R[D]$ for different species of $D$ mesons are
collected in  Table~\ref{tab0}.   The E687 measurements of $\alpha[D^+]$ as a
function of $E_\gamma$, $p_\perp^2$ and $x_F$ are shown in Figs.~\ref{Eg},
\ref{ptsqr} and \ref{xf}, respectively.  The E687 results are not corrected for the
acceptance from a cut on the number of charged tracks at the photon interaction
vertex, because the E687 hadronization model introduced a strong correlation
between the acceptance and the asymmetries. This introduces an unknown systematic
error in the comparison with theory. For both E687 and E691, we have combined the
results from  the 2 different decay channels of $D^{*+}$ into single numbers. The
E687  result for the 2 different decay channels for $D^0$ can be combined into the
single  number $R=1.035\pm0.021$. Unfortunately, the $D^0$  sample from E687 was
subjected to a  ``no $D^*$ tag" requirement that reduced the feeddown from
$D^{*+}\rightarrow D^0$  to about $20\%$  of the total without reducing the
feeddown from  $D^{*0}$. We have attempted to undo the effects of that requirement,
which only complicates the comparison with theory. A naive prediction for the
fraction of $D^0$'s from $D^{*+}$ feeddown can be obtained by neglecting
heavy-quark recombination and using isospin and heavy-quark spin symmetry in Eq.~(\ref{D0}):
$3\cdot 0.677/(1+3\cdot 0.677+3)=34\%$. A feeddown of $20\%$ can be obtained by
reducing the fraction multiplying $\sigma[D^{*+}]$ in Eq.~(\ref{D0}) to 0.333:
$3\cdot 0.333/(1+3\cdot 0.333+3) = 20\%$.  The inclusive $D^0$ events should
therefore consist of  approximately $(1+3\cdot 0.333+3)/(1+3\cdot 0.677+3) = 83\%$
$D^0$ with no $D^*$ tag events and $17\%$  $D^{*+}$ events.  Since the observed
values of $\alpha$ are small, the value of $\alpha[D^0]$ is simply given by the
appropriately weighted values of  $\alpha[D^0, {\rm no}\; D^*\;{\rm tag}]$  and
$\alpha [D^{*+}]$, up to corrections of order $\alpha^2$. We therefore take the
E687 result for the asymmetry variable $\alpha$  for $D^0$ to be
\beq
\alpha[D^0] = 0.83\,\alpha [D^0, {\rm no}\; D^*\;{\rm tag}]
+ 0.17\,\alpha [D^{*+}]\,.
\eeq
We obtain $\alpha[D^0] = -0.023\pm 0.009$, which corresponds to the value of $R$
listed in Table~\ref{tab0}.

The E687 experiment has a broad-band photon beam with average energy $\langle
E_\gamma \rangle = 200\, {\rm GeV}$. To calculate the  $p_\perp^2$ and $x_F$
distributions, theoretical cross sections were convoluted with the photon beam
spectrum of the E687 collaboration.\footnote{A convenient
parametrization of this spectrum can be found in  Section 5 of
\cite{Frixione:1994dg}.} Our results are nearly identical if instead of using the
real spectra, a monochromatic 200 GeV beam is used instead. For the  $E_\gamma$
distribution, the data correspond to events collected in energy bins  centered
around 40, 120, 200 and 280 GeV.  For simplicity, we calculated each data point
with fixed photon energies corresponding to the center of each bin.  Feynman $x_F$
is defined by
\beq
x_F= {p_z \over |p_z|_{\rm max}}\approx
{{m_\perp \rm sinh}(y) \over \sqrt{s_{\gamma N}}}\, ,
\eeq
where $m_\perp^2 = p_\perp^2 +m_c^2$, $y$ is the rapidity of the $D$ in the
photon-nucleon center-of-mass frame and $s_{\gamma N}$ is the photon-nucleon
center-of-mass energy squared. The positive $z$-axis is defined by the incident
photon direction. We use the approximate expression  for $x_F$ in our
calculations.  The experimental results were all subject to the cut $x_F >0$.

To fix the parameters $\rho_{\rm sm}$ and $\rho_{\rm sf}$, we did a least-squares
fit to the E687 data consisting of the ratios $R[D]$ for the four mesons in Table
\ref{tab0}, the $p_\perp^2$ distribution for $\alpha[D^+]$ consisting of the five
data points in Fig.~\ref{ptsqr}, and the $x_F$ distribution for $\alpha[D^+]$
consisting of the first four data points in Fig.~\ref{xf}. Data from the last $x_F$
bin was not included in the fit because the theory is expected to be less reliable
near $x_F = 1$. In this corner of phase space, one expects increased uncertainty
due to higher order corrections, nonperturbative fragmentation effects, the
difference between the charm quark mass and $D$ meson mass, etc.  The results of a
two parameter fit gives a value of $\rho_{\rm sf}$ that is small and negative.
Since $\rho_{\rm sf}$ is proportional to a fragmentation  probability, this
negative value is unphysical. Therefore, we set $\rho_{\rm sf} = 0$, and the fit
then yielded $\rho_{\rm sm} = 0.15$.

Theoretical predictions for the kinematic dependence of $\alpha[D^+]$ are compared
to the E687 results in Figs. \ref{Eg}, \ref{ptsqr} and \ref{xf}.   Heavy-quark
recombination describes the observed asymmetries well. In particular, the
$E_\gamma$ dependence and the $p_\perp^2$ distribution are fit very well. The
agreement with the $x_F$ distribution is not as good, but the largest discrepancy
is about 2.5 $\sigma$ in the last $x_F$ bin, where theoretical errors are
greatest.  Predictions for the $R[D]$ ratios using the parameters $\rho_{\rm sm}=
0.15$ and $\rho_{\rm sf}=0$ are compared to the E687 and E691 data in
Table~\ref{tab0}.  Most predictions for the $R[D]$ ratios are within 1 $\sigma$ of
the data. However, the predictions for the $R[D^{*+}]$ ratios of both experiments
are below the data by about 2.6 $\sigma$. A better fit to the $R[D]$ ratios could
be obtained if the larger value $\rho_{\rm sm}=0.3$ is chosen, but at the expense
of a worse fit to the kinematic distributions for $\alpha[D^+]$.

The heavy-quark recombination mechanism reproduces some interesting qualitative features of the
data. One feature is that the asymmetries decrease as $E_\gamma$ increases. There
are several effects that all tend to decrease the asymmetry. First, the ratio of
quarks to gluons at small $x$ decreases with energy. Furthermore, the ratio of $q$
to $\overline{q}$ at small $x$ approaches 1, further diluting the asymmetry.
Finally, the  parton cross sections for heavy-quark recombination fall faster than
photon-gluon fusion as the parton center-of-mass energy grows.  A second
qualitative feature is that $R[D^+],R[D^{*+}]>R[D^0]$.
This is largely because the
recombination cross section for $\overline{c} d$ is greater than that for
$\overline{c} u$. Because the beryllium nucleus is very close to an isosinglet
target, the difference in these cross sections is not due to the parton
distribution functions, but rather to the parton cross sections in Eq.~(\ref{ss})
and Eq.~(\ref{st}).  The differences appear because the coupling to the photon does
not respect isospin symmetry: $\kappa = 1$  for $\overline{c} u$ and $\kappa =
-{1\over 2}$  for $\overline{c} d$.

\begin{figure}[!t]
  \centerline{\epsfysize=14.0truecm
  \begin{turn}{270}
    \epsfbox{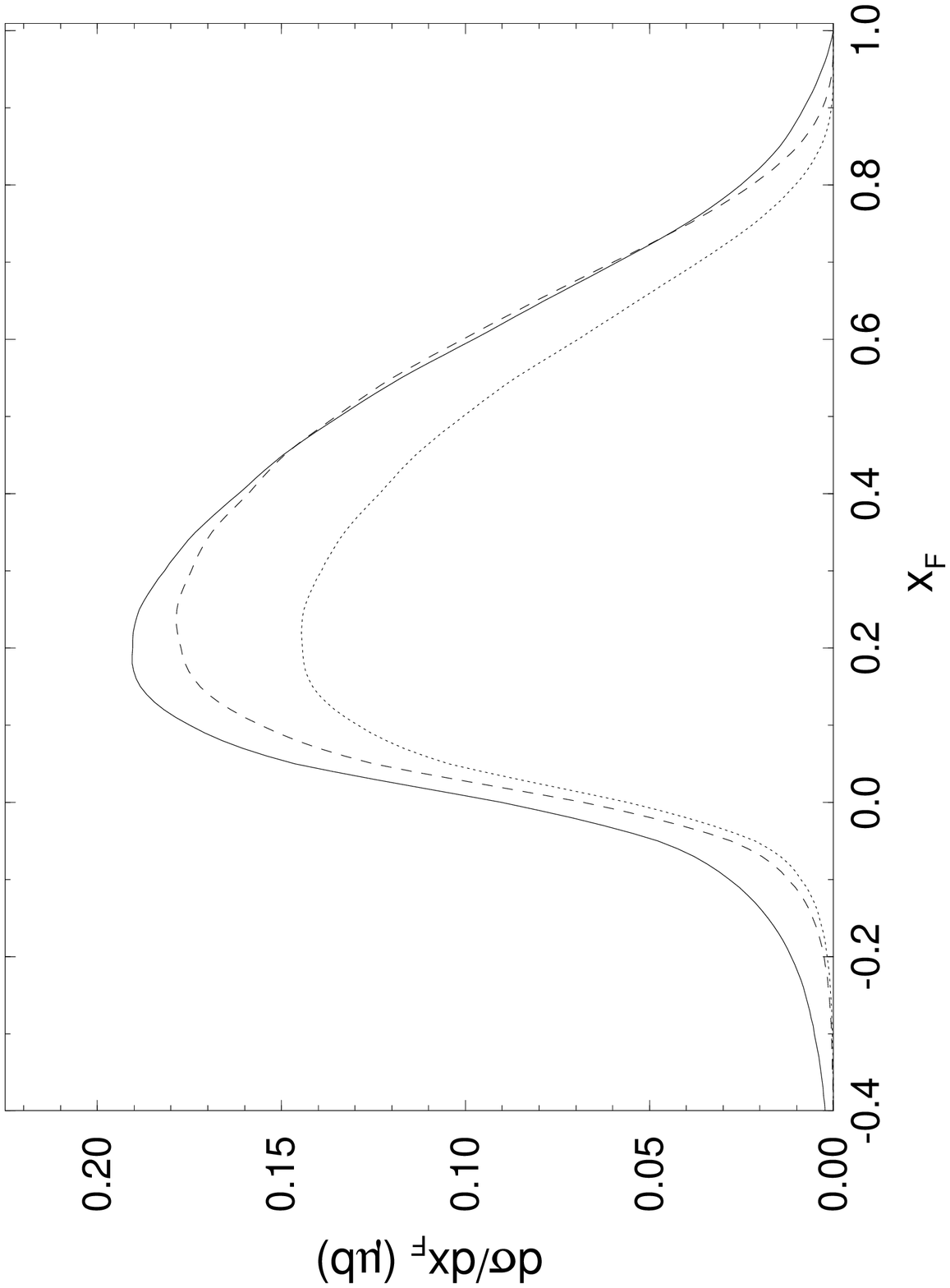}
  \end{turn}  }
  \vspace{0.25 in}
{\tighten
\caption{Inclusive cross sections $d\sigma/dx_F$ for $D^-$ (solid curve) and
$D^+$ (dashed) with $\rho_{\rm sm}=0.15$, $\rho_{\rm sf}=0$. The dotted line shows
the leading order photon-gluon fusion contribution to the $D^{\pm}$ cross sections.}
\label{dsdxf}}
\end{figure}

In Fig.~\ref{dsdxf}, the predicted inclusive cross sections, $d\sigma/dx_F$, for
$D^+$ and $D^-$ mesons are shown.  Also shown is the contribution to the cross
section from the leading order photon-gluon fusion process. Again we have used the
E687 photon beam spectrum with $\langle E_\gamma \rangle = 200\,{\rm GeV}$. For
$\rho_{\rm sm}= 0.15$ and $\rho_{\rm sf}=0$,
heavy-quark recombination contributes about 25\%
of the total cross section for both the $D^+$ and $D^-$. The cross sections are
small for $x_F <0$, however, in this region the $D^-$ cross section is much greater
than the $D^+$ cross section. Thus, heavy-quark recombination predicts that $|\alpha[D^+]|$ is
largest in the region of phase space excluded by the experimental cut $x_F>0$.
This is because recombination is most important when the $(\overline{c}q)$
emerges in the forward direction of the light quark in the nucleon,
as shown in Eq.~(\ref{ratio}).

The calculations of this paper neglected NLO corrections to the leading order
photon-gluon fusion diagrams. If the NLO corrections are incorporated by including
a K factor, the extracted value of $\rho_{\rm sm}$ must be multiplied by the same
K factor to obtain the same asymmetry.  As stated earlier, there is at least a 30\%
uncertainty in this parameter due to $1/m_Q$, $SU(3)$ breaking and $1/N_c$
corrections, and a similar size uncertainty in $\rho_{\rm sf}$.

\section{Other applications}

The heavy-quark recombination mechanism also can be used to calculate  the
asymmetries of charmed baryons such as the $\Lambda_c^+$. Although the
$\overline{c}q$ recombination mechanism will contribute at some level, the most
important contribution is expected to come from $cq$ recombination. We expect the
asymmetry to be dominated by the analog of process $a)$ in Eq.~(\ref{direct}),
which would lead to a positive asymmetry $\alpha[\Lambda_c^+]$. In the measurements
from E687 and E691 in Table I, the central values correspond to a positive
asymmetry, but the errors are large enough that the asymmetry is not statistically
significant. The FOCUS experiment at Fermilab will measure the asymmetries for
$\Lambda_c^+$ with much higher statistics and will also measure the asymmetries for
other charmed baryons, such as $\Sigma_c^0$, $\Sigma_c^{++}$  and
$\Sigma_c^{*++}$.  It will be interesting to compare these measurements with the
predictions of our heavy-quark recombination mechanism.
Quantitative predictions
will involve nonperturbative factors of the form $\rho[(cq)^n \rightarrow
\Lambda_c^+]$. Once these are fit to reproduce the overall asymmetries, the
kinematic distributions of the asymmetries will be predicted.

The heavy-quark recombination cross sections calculated in this paper
predict that $\alpha[D_s^+]$ should have the opposite sign as
$\alpha[D^+],\alpha[D^0]$.  However, it is important to point out
that the $D_s^+ - D_s^-$ asymmetry is generated by the fragmentation
of charm quarks from process $b)$ in Eq.~(\ref{backside})
in which the charm antiquark is involved in the recombination process.
In $cq$ recombination, there is a process analogous to process $b)$
which can dilute the asymmetry generated by
$\overline{c}q$ recombination. To provide reliable predictions for
$\alpha[D_s^+]$ asymmetries,
both $\overline{c}q$ recombination into mesons
and $cq$ recombination into baryons need to be included.

Recent photoproduction measurements by the ZEUS detector
at HERA seem to indicate that NLO calculations
underestimate the $D^*$ production rate in the forward proton
direction \cite{ZEUS}, although
this effect has not been confirmed by the H1 detector \cite{Breitweg:1999yt}.
If there is an excess of $D^*$ in the forward direction,
our heavy-quark recombination mechanism may be able to explain it.
Heavy-quark recombination effects should be most important
in the forward proton direction, because the ratio  of the partonic cross sections
for $\overline{c}q$ recombination and photon-gluon fusion is highest in this
direction. The forward nucleon direction is excluded in fixed target
photoproduction measurements by the $x_F>0$ cut, but it is experimentally
accessible at the HERA $ep$ collider. For quantitative calculations of the charm
asymmetries at HERA energies, it will be necessary to take into account the
contribution of resolved photons.

A large charm asymmetry has been observed in fixed-target hadroproduction
experiments, where it is called the
``leading particle effect"\cite{Aitala:1996hf}:
charmed hadrons sharing a valence quark in common with the beam hadron
are produced
more copiously than other charmed hadrons in the forward region.
Our heavy-quark recombination
mechanism can be used to predict these asymmetries.
The relevant parton processes are the same as those considered
in \cite{Braaten:2001bf} for $B$ meson production
at the Tevatron. Because fixed-target hadroproduction involves lower energies,
heavy-quark recombination is expected to play a more important role
in these experiments.    It would be interesting to see
if heavy-quark recombination could provide a quantitative
description of the leading particle effect.

We thank  P. Garbincius, R. Gardner and J. Wiss for providing useful information
about the E687 experiment.  This work was supported  by the Department of Energy
under Grant No.\ DE-FG02-91-ER4069 and by the National Science Foundation
under  Grant No.\ PHY--9800964.

\newpage

\begin{table}
{\tighten  \caption{Anticharm/charm production ratio $R= N_{\overline D}/N_D$:
experimental results from E687 and E691, and theory
predictions with $\rho_{\rm sm}=0.15, \rho_{\rm sf}=0$.  }
\label{tab0} }
\begin{center} \begin{tabular}{ccc|ccc|ccc} 
&  Particle & & E687~\cite{Frabetti:1996vi}&  Theory & &
E691~\cite{Anjos:1989bz} & Theory
  \\
&  & &  $\langle E_\gamma \rangle = 200\:{\rm GeV}$  &   &
& $\langle E_\gamma \rangle = 145\:{\rm GeV}$   &  &   \\
 \hline
& $D^+ $   & &  $1.08\pm0.02$    & $1.06$  && $1.04\pm0.03$ &$1.07$&\\
& $D^0$  &  & $1.047\pm0.018$  &   $1.03$   && $1.08\pm0.03$ & $1.03$&\\
& $D^{*+}$ & & $1.112\pm0.024$  &  $1.05$  && $1.190\pm0.049$ &$1.06$ &\\
& $D_s^+$  &  & $0.95\pm0.10$ &   $0.91$   && $0.92\pm0.14$ & $0.88$ &\\
& $\Lambda_c^+$   & & $0.93\pm0.14$    & - &&   $0.79\pm0.17$ & -& \\ \hline
&  $D^{*0}$ & & - &  $1.02$  && - & $1.02$ &\\
&  $ D_s^{*+} $ & & -  &  $0.91$  &&- & 0.88& \\
\end{tabular} \end{center}
\end{table}

\newpage

\end{document}